# Nambu at Work

Peter G.O. Freund

Enrico Fermi Institute and Department of Physics
University of Chicago, Chicago, IL 60637 USA

**Abstract**:

This paper is based on a talk delivered on 16 November, 2015 in Osaka at the *Nambu's Century: International Symposium on Yoichiro Nambu's Physics.*

Yoichiro Nambu, whose life and seminal contributions to Physics we celebrate here, went in 1952 to the Institute for Advanced Study in Princeton. Shortly after his arrival there, J. Robert Oppenheimer, the Institute's director, put Yoichiro and the other new arrivals on notice that though Albert Einstein was a professor at the Institute, and therefore had an office there, nobody was to disturb the great man without first receiving special permission personally from Oppie. Most people would spend a year or two in the same building with Einstein and then spend a whole lifetime regretting not to have met him. Yoichiro decided that he will meet Einstein, no matter what Oppie says. He knew Bruria Kaufmann, Einstein's assistant at that time, and with her help got to visit the great physicist. Einstein was very friendly and visibly happy that finally one of the young people had bothered to visit him. Einstein asked Yoichiro what was going on in particle physics, and was rather skeptical about separate nucleon and meson fields for which he saw no deeper reason.

It is well-known that Einstein never learned to drive. Yoichiro offered him a ride, ran to his car to open the door, and from the driver's seat snapped a picture of Einstein, it's composition worthy of a major photographer.

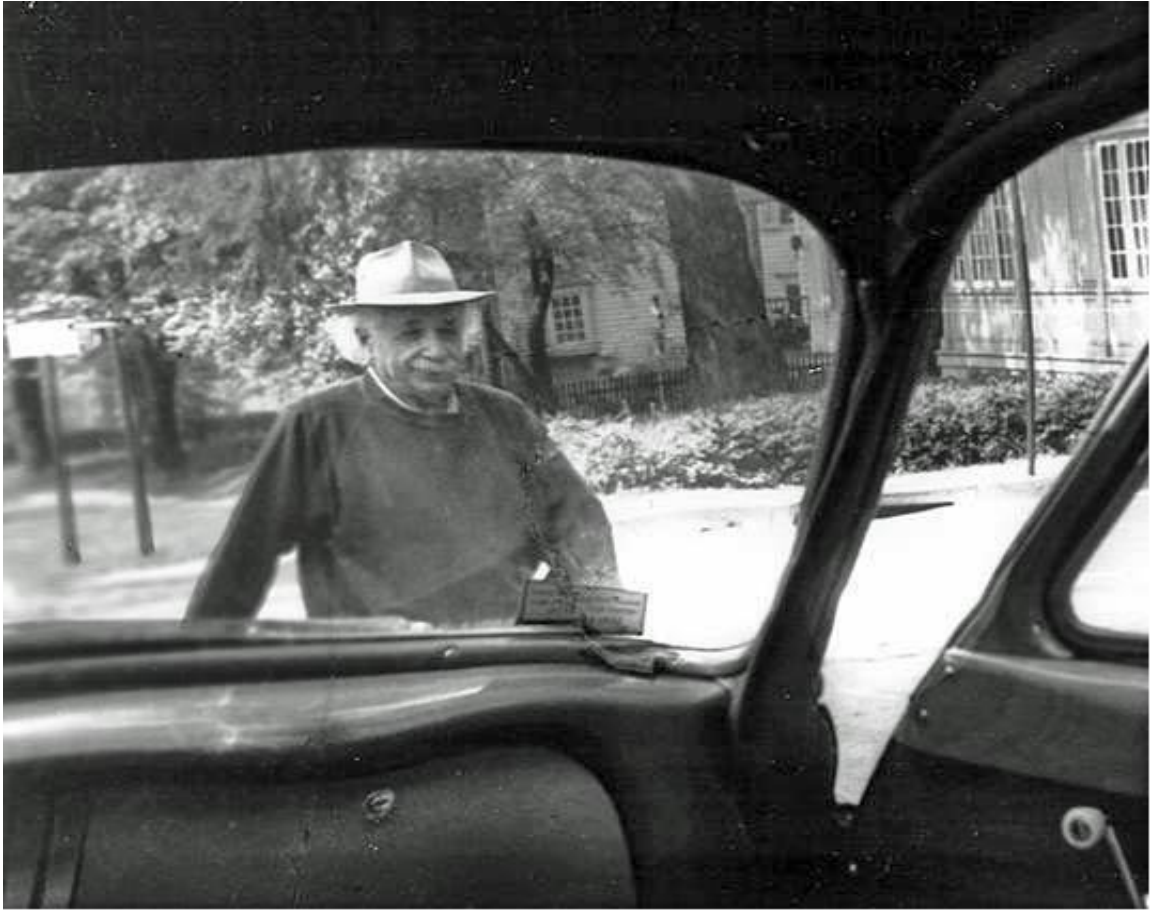

Nambu offered Einstein a ride in his car and took this picture of him

I brought this episode up at the very beginning of this talk, because there is a certain resemblance between the way the major works of Einstein and Nambu were conceived. In the construction of a relativistic theory of gravitation, Einstein used as a guide Mach's principle [1]. Actually this principle was not really embedded in general relativity, the end-product of Einstein's work. Though many people [2] have tried since, to this day Mach's principle is not incorporated in the theory, which can be understood without any reference to this beautiful somewhat vague and philosophical principle.

When constructing with Moo-Young Han [3] the color gauge theory of strong interactions Yoichiro, as he told me, was also guided by a philosophical principle, the quasi-Hegelian so-called "three stages" principle of Mitsuo Taketani [4]. The physics of mesons and baryons was for strong interactions the first *phenomenological* stage, somewhat like the 19th century spectroscopic results were for quantum theory. Similarly the Gell-Mann-Zweig quarks [5] represented the second *substantive* stage, like Bohr's atom, but the deeper *essentialist* stage, Taketani's third and final stage, the full grown theory like Heisenberg and Schrödinger's quantum mechanics, was to involve something unexpected in the form of a non-abelian color SU(3) gauge theory. Yoichiro did not sympathize with Taketani's politics, but, though it may not be well-known, he subscribed to Taketani's three-stages philosophy for physics theories, and was convinced that it was useful to him in the color gauge work, and as we shall see, in his string theory work as well.

The idea of a gauge theory as the basis for strong interactions was in the back of Yoichiro's mind for quite some time. Before his proposal of the non-abelian color gauge theory, Yoichiro had tried to gauge the Pauli-Pürsey-Gürsey symmetry [6], which mixes spinorial Fermi fields with their charge conjugates, and which had played a role in Heisenberg's much heralded but flawed unified theory [7]. I remember a number of conversations on this topic. That did not lead to any interesting results, but color gauge theory did.

Heisenberg's unified theory also influenced Yoichiro's discovery of spontaneous symmetry breaking. Besides its origin in Nambu's work on BCS [8] theory, the four-Fermi interaction starting point of the Nambu-Jona-Lasinio papers [9] reflects also the influence of Heisenberg. In fact, Heisenberg has thought of breaking the symmetries of his four-Fermi lagrangian and made an inspired analogy with ferromagnetism [7], but he missed the crucial feature of such a symmetry breaking mechanism: the appearance of a Nambu-Goldstone boson, in this case the so-called spin-wave or magnon, discovered by Heisenberg's former student Felix Bloch, and further studied in an important paper by Holstein and Primakoff [10], though the wide generality of this idea was not recognized by any of them.

It is amusing that the four-Fermi interaction in the BCS theory of superconductivity, with the electron as the fermion, is what gave that theory its "from basic principles" cachet. The bosonic Ginzburg-Landau (GL) theory [11] was viewed by the solid-state-physics community as nothing more than a phenomenological model, anyway until Gor'kov showed [12] that GL can be derived from BCS.

In particle physics the situation is reversed. Shortly after Nambu's fundamental papers, in which the degeneracy of the vacuum and the appearance of massless bosons was discovered, Goldstone (giving full credit to Nambu) recast [13] their approach in an elegant bosonic classic field theory language, which became the preferred version of the particle physics community. Goldstone then proved his brilliant and far-reaching theorem [14], which governs the appearance and number of the massless particles. These are the reasons why these massless particles became known as "Goldstone bosons," and only in recent times were given by the particle physics community the more appropriate name of Nambu-Goldstone bosons.

For decades, in Chicago, at the weekly Enrico Fermi Institute theory seminar, some young physicist freshly out of graduate school would give a talk about "Goldstone bosons" with Nambu sitting in the first row, and not batting an eye. The beauty of all this is that in the long run, the history of physics always sorts itself out in a fair way.

When gauge fields were added to the mix, this led to the beautiful Englert-Brout -Higgs [15] phenomenon, and here again the earlier Englert-Brout approach was phrased in a quantum field theoretic language, whereas Higgs' version used Goldstone's classic field theory language. Again, the particle physics community opted for this simpler version and named the outcome the Higgs phenomenon, and again only in recent times has the more appropriate name of Englert-Brout-Higgs phenomenon been generally adopted.

In those days before the internet and even the Xerox copier, Englert and Brout mailed a copy of their preprint to Yoichiro, who liked it and came to my office to discuss it. We, of course, wanted to see whether this mechanism could be used to give mass to the by then well-known vector mesons, assuming that they were the gauge bosons of a Sakurai-type [16] SU(3)-flavor-gauge theory of strong interactions. By lunchtime, we had convinced ourselves that if this was the mechanism by which these vector mesons acquired their mass, then the K* had to be much heavier than the ρ, and only the charged ρ's were to be massive, while their neutral partner would stay massless. This was all seriously at odds with the experimental situation, and we concluded that, though it was a pity, this mechanism was of no use. It remained to be discovered by Weinberg [17] and Salam [18], that this mechanism was designed not for giving mass to some hadrons, but for the higher task of giving mass to the carrier bosons of the weak interactions.

Now let me move forward in time to the late sixties, the era of the birth of string theory. Again a phenomenological stage, the so-called two-component duality [19], started the game, the two components being destined to correspond to the open and closed strings. It was followed by the substantialistic stage of the Veneziano model [20], and in the hands of Nambu [21], Susskind [22] and Nielsen [23] it took on the *essentialistic* form of string theory.

I remember one morning in Yoichiro's office, Yoichiro telling me about his insight. We often tested our ideas on each other, and I must mention that quite a few of my ideas became much clearer after they were "tested" on Yoichiro. To appreciate Yoichiro's presentation of his new results one had to know him and be familiar with his personality. He would start by mentioning some very well-known results and putting on the blackboard in his beautiful handwriting a list of famous names. If you did not know Yoichiro, all this had the effect of making you think that he was going to tell you about a small detail in some well-known previous theory, a detail which maybe wasn't even new. But then, after about ten minutes or so, suddenly a new and often very deep approach was coming to life in front of your eyes. It was easy to understand what he was saying, but where all this beautiful new stuff was coming from, remained wrapped in total mystery.

As one brief example, when he arrived at string theory, what took him by quite some surprise was the exponential vertex operator, the normal-ordered version **:exp[ik X(0,τ)]:**. He came up with a very simple argument for this vertex: If I want to carry out a space translation by **Δx**, I use the familiar operator **exp(iΔxP)**, with **P** the momentum operator, the generator of space translations. But here I want to modify not the position by **Δx**, but rather the momentum of the state, by **k**. So, by analogy, I use the operator **:exp[ik X(0,τ)]:** with **X(0,τ)** generating the required momentum "translation" during an interaction. Justifiably, he was proud of this argument. After all, this Nambu vertex turned out to be a mathematically very useful and richly endowed object.

There are two styles of doing seminal work in theoretical physics. In one style, an extremely clever physicist familiar with current theories and experimental results realizes that there is something missing for all this to gel and comes up with the theory that does the trick. This is the style of Einstein, of Heisenberg, of Yukawa and of Gell-Mann. Anybody can understand how they got to their theories. You can do both, admire and use the end result, *and* understand perfectly well where their ideas came from.

In the other style, the extremely clever physicist realizes that something is missing for a theory to attain its deeper meaning, and then supplies this missing element. This is the style of Einstein, of Dirac, of Feynman and of Nambu. You can again admire and use their end results, but how they ever got these ideas remains, as I said, wrapped in mystery. Yes, Einstein did work in both styles.

At this point it may be appropriate to say a few words about Nambu the man. Beside his marvelous modesty, Yoichiro was fully attuned to the American ways, while still hanging on to his Japanese customs. In Japan politeness and conflict avoidance dictate a minimal use of the word "no." In Yoichiro's case this meant no use at all of this word. No matter how preposterous the request, after a pause he would always answer with a "yes," but the length of the pause measured his wish to say "no," the longer the pause, the more negative the response. Yoichiro's definition of the word "no" is a "yes" spoken after an infinitely long pause.

This led to some funny situations when Yoichiro became chairman of our Physics Department. It also led to one of Yoichiro's doctoral students staying around for many years and when the student did not leave on his own, Yoichiro let him graduate, simply because he could not bring himself to tell the student that he was *not* fit for doing theoretical physics.

I should mention here that Yoichiro had a number of excellent students: Lou Clavelli, Sumit Das, Savas Dimopoulos, Tony Gherghetta, Markus Luty, Burt Ovrut, Richard Prange, Jorge Willemsen, Motohiko Yoshimura, to name but a few.

Of the many brilliant scientists I met over the years, Yoichiro is one of the very few about whom I am certain that he was a genius.

**Appendix**

It may be somewhat unusual to find an appendix in an article of this kind, but I would like to reproduce here part of an email message I got from Yoichiro on 5 December, 2013. That message starts with some comments on Japanese politics, which I shall omit here, and then goes on to a science question:

*"Dear Peter*

*............*

*By the way I would like to ask you a question. In the course of studying Bode`s law I found a mysterious paper on a derivation of the Schroedinger equation from classical dynamics by regarding time evolution not as a simple time derivative but as a stochastic process.   Nelson E., Derivation of the Schrödinger equation from Newtonian mechanics, Physical Review, Vol. 150, No. 4, 1079-1084 (1966).*
*I have not understood the paper yet. (There are careless errors in equations.) Did you know about it?*
*  YN"*

For completeness, here is my answer:

*"Dear Yoichiro,*

*Yes, I have seen Nelson's derivation of the Schroedinger equation. I always thought of it as a kind of perverse vindication of Hamilton's and Lagrange's work. What I mean to say is that there are three approaches to classical mechanics, Newton's, Hamilton's and Lagrange's. The latter two, though originally developed as pure mathematical physics, contained the remarkable optics-mechanics analogy, and naturally led to the Heisenberg, the Schroedinger and the Feynman approaches to QM. It would be hard to see how QM would have been discovered, had Hamilton and Lagrange not done their work in the nineteenth century. But that leaves open the question as to whether QM could have been arrived at, had Newton's approach been the only one known. Nelson answered that question in the affirmative, but I doubt that QM would have been discovered as "naturally" and as early, had the Newton/Nelson approach been the only one available. I find Nelson's work to be more of the "for completeness' sake" type, although I admit that it may still provide useful clues in the future.*
*Best*
*Peter"*

Bode's law---or more accurately the Titius-Bode law---mentioned here, states that in our solar system, in astronomical units, the semi-major axes of the planets' orbits are given by the formula

$$a(n) = 0.4 + 0.3 \cdot 2^n \qquad \text{with } n = -\infty, 0, 1, 2, 3, 4, ....$$

so that $n = 1$, $a(1) = 1$ corresponds to the earth, as befits astronomical units. This law was arrived at in the eighteenth century and it is obeyed to within an error of 5% by the first eight planets. For Neptune the law's prediction is too large by some 30%, and for Pluto it is too large by almost a factor 2, but improved versions of the law have been proposed, which accommodate Neptune and Pluto, and are valid for other planetary or satellite systems as well. The relevance of a Schrödinger-type equation for such systems has been observed by Albeverio, Blanchard, Høegh-Krohn [24] and others. These issues are reviewed and expanded in a paper by Scardigli [25]. In the simplest way, the appearance of a Schrödinger-type equation can be seen as follows [25]. In a Bohr-like model let us require the usual equality of the gravitational attraction and centrifugal forces,

$$GMm/r^2 = mv^2/r \qquad (1)$$

and then impose the new type of "quantization" condition

$$J/m = vr = se^{\lambda n} \qquad (2)$$

instead of the Bohr quantization

$$J = \hbar n. \qquad (3)$$

Here $M$ is the sun's mass, $m$ is the mass of the planet, which on account of the quantization condition (2) and of the equivalence principle which guarantees the appearance of the same mass $m$ on both sides of Eq. {1}, ultimately cancels out, $v$ is the planet's velocity, $r$ its distance from the sun, $G$ is Newton's constant, $J$ is the orbital angular momentum whereas s and λ are new parameters. Equations (1) and (2) yield a law with *n* in the exponent,

$$r_n = r_0 e^{2\lambda n} \text{ with } r_0 = s^2/GM. \qquad (4)$$

which for large *n* and $e^{2\lambda} > 1$ produces a Titius –Bode type law. Comparing with experimental data on various planetary or satellite systems, the parameter *s*, unlike Planck's constant which it replaces, does not take a universal value, but changes from system to system, However the parameter $\lambda$ remains essentially the same [25].

Finally, just like Bohr's atom to make sense, requires the well-known Schrödinger equation, so this Titius-Bode type system, with its different angular momentum operator, yields a different Schrödinger-like equation [24, 25]. There are of course no quantum jumps from one level to another, no superposition principle, etc.... The stochastic feature of this Schrödinger-type equation originates in the classical chaotic dissipative processes which, over a sufficiently long time, bring the original dust in the proto-planetary system to stabilize in the Titius-Bode orbits. At a deeper level these ideas are connected to the "determinism beneath quantum mechanics" advocated by 't Hooft [26]. Given these facts, I wonder where Yoichiro was headed when pursuing these ideas. As always, we are probably going to find out ten years from now why he was on the right path already in 2013.